\documentclass[fleqn]{wlscirep}
\usepackage[utf8]{inputenc}
\usepackage[T1]{fontenc}
\usepackage{subcaption}         %
\usepackage{caption}
\usepackage{enumitem}
\usepackage{lipsum}
\usepackage{siunitx}
\usepackage{multirow}
\usepackage{bbm}

\usepackage{xr}
\title{Graph-theoretic design of lasing networks for physical vision}

\author[1,2,$\dagger$]{Paul Obernolte}
\author[1]{Jakub Dranczewski}
\author[1]{Yixiu Yin}
\author[1]{Tobias Farchy}
\author[1]{Wai Kit Ng}
\author[1,2]{Tobias Simonsen}
\author[1,2]{Elias Großhauser}
\author[3]{Alexis Arnaudon}
\author[4]{Robert L. Peach}
\author[5]{Mauricio Barahona}
\author[1, $\dagger$]{Riccardo Sapienza}
\author[1,6,7,$\dagger$]{Jack C. Gartside}
\author[1,5,$\dagger$]{T. V. Raziman}

\affil[1]{Blackett Laboratory, Department of Physics, Imperial College London, London, United Kingdom}
\affil[2]{Heidelberg University, Heidelberg, Germany}
\affil[3]{Blue Brain Project, École Polytechnique Fédérale de Lausanne (EPFL), Campus Biotech, Geneva, Switzerland} %
\affil[4]{Department of Neurology, University Hospital Würzburg, Würzburg, Germany} %
\affil[5]{Department of Mathematics, Imperial College London, London, United Kingdom}
\affil[6]{London Centre for Nanotechnology, Imperial College London, London, United Kingdom}
\affil[7]{Institute for Materials Research, Tohoku University, Sendai, Japan}
\affil[$\dagger$]{email: p.obernolte@thphys.uni-heidelberg.de, r.sapienza@imperial.ac.uk, j.carter-gartside13@imperial.ac.uk, t.raziman@imperial.ac.uk}

\begin{abstract}
Physical neural networks perform learning through the intrinsic nonlinear dynamics of matter.
Optimising their design presents a considerable challenge: complex many-body physics can provide powerful computation, but are expensive to simulate and large experimental optimisation runs are impractical to fabricate.
Hence, the high-dimensional space of possible network topologies cannot be effectively directly searched.
Here, we show that this search can be efficiently performed in an abstract graph space that is vastly cheaper to explore.
Using random lasing networks -- composed of interconnected nanoscale waveguides and hosting strongly coupled lasing modes -- as an exemplar physical vision system, we establish a quantitative three-layer link: simple graph-theoretic metrics predict the nonlinear lasing physics, which in turn predicts vision performance.
After validating this relationship using physical simulations, we exploit it to drive an evolutionary algorithm using graph metrics, producing network topologies that outperform random designs at a fraction of the computational cost (3000$\times$ speed-up compared to physical simulation).
On simulated image-classification tasks, graph-optimised networks substantially improve classification accuracy.
As our framework operates on network topology rather than substrate-specific physics, we anticipate it can transfer to other network-based physical learning systems, providing an efficient route for the directed design and optimisation of complex, strongly-interacting physical neural networks.

\end{abstract}

\begin{document}

\flushbottom
\maketitle
\thispagestyle{empty}

\section*{Introduction}

Physical learning systems exploit the dynamics of matter itself to perform computation\cite{tang_bridging_2019,markovic_physics_2020,schuman2022opportunities,li_photonics_2025,momeni2025training,kudithipudi2025neuromorphic,brunner2025roadmap,kurebayashi2026metrics}, but this creates a central design problem: the same complex many-body nonlinear interactions which make these systems computationally rich also make them challenging to optimise.
The origin of the many-body interactions divides physical learning systems into two classes:
The first class is built from discrete functional units interconnected electronically or digitally, such as magnetic tunnel junctions\cite{ross2023multilayer,duffee2025integrated,menshawy2026remotely}, memristors\cite{yao2020fully,jebali2024powering,renaudineau2026forward}, and reconfigurable nonlinear processors\cite{taglietti2026learning,escudero2026physical,zolfagharinejad2025analogue}.
The latter class implements network connectivity and parallel processing directly in the physical interactions, such as in strongly interacting nanomagnetic metamaterials\cite{gartside2022reconfigurable,vidamour2023reconfigurable,stenning2024neuromorphic} and nonlinear photonic lasing systems\cite{skalli2022computational,skalli2022photonic,ng2025fewshotretinomorphicvisionnonlinear,skalli2025model}.
These physically-coupled approaches reduce digital/electronic overheads and the complex nonlinear coupling dynamics provide rich computational resources, but their complexity raises significant challenges for network design.
In network-based physical learning systems, performance is not defined solely by a list of individual device parameters, but by the topology of the underlying network which governs physical interactions and information flow.
Exhaustively simulating, fabricating, and testing candidate topologies is impractical and prohibitively expensive.
A useful design theory must identify low-cost metric descriptors that predict the emergent physical dynamics responsible for computation.

Random lasing networks\cite{saxena2025designed,saxenaSensitivitySpectralControl2022,gaio2019nanophotonic,sapienza2022controlling,sapienza2019determining,cao2005random,cao2003lasing,wiersma2008physics,ng2025fewshotretinomorphicvisionnonlinear} provide a stringent test of this problem.
Their multimode lasing spectra depend sensitively on waveguide connectivity, spatial mode overlap, and nonlinear gain competition\cite{saxenaSensitivitySpectralControl2022}.
These properties can produce emergent spatially-sensitive nonlinear responses naturally suited for neuromorphic vision including feature detection, classification, and segmentation, particularly when training data is limited\cite{ng2025fewshotretinomorphicvisionnonlinear}.
The complexity of these dynamics makes lasing networks highly expensive to simulate and challenging to optimise directly. 
This raises a general question: can the useful dynamics of a physical learning network be predicted and tuned from the network graph topology alone?

Tools developed for graph theory are well suited to answering this question, and have been used to interpret and design systems in neuromorphic computing\cite{loeffler_topological_2020} and other fields\cite{bindgen_connecting_2020,lawrie_application_2024,maurizi_designing_2025}.
However, these works optimise passive structural or mechanical responses - whether abstract graph properties can predict the emergent nonlinear dynamics and learning performance of an active, strongly-interacting physical network, and drive efficient optimisation and design, has not been shown.
If possible, such an approach would allow large design spaces to be screened cheaply, identifying promising candidate networks for later validation by detailed simulation, fabrication, and experiment.
It could also reveal which network properties actually drive performance in complex physical systems.

Here, we introduce a graph-theoretic framework for the efficient design and optimisation of network-based physical systems, focusing on random lasing networks as a model platform.
We show that simple graph metrics, including total edge length, clique size, and edge betweenness centrality, correlate with key physical properties of the lasing network, including the number of lasing modes and the degree of inter-mode competition - previously shown to be a key driver of vision performance in lasing networks\cite{ng2025fewshotretinomorphicvisionnonlinear}.
These physical properties correlate with proxy metrics for vision performance/physical machine learning, including excitatory/inhibitory mode responses and the number of distinct image features detected by the network.
We then use these graph-based metrics to guide an evolutionary optimisation algorithm, producing network topologies that outperform large populations of randomly generated designs.
When evaluated on simulated image-classification tasks, these optimised networks show substantially reduced classification error compared to networks with poor graph-based metrics - by 2$\times$ between the best and worst networks considered for the Pneumonia chest X-rays task, and by 1.5$\times$ for two-class K-MNIST.

These results demonstrate the promise of moving network design and optimisation into computationally-efficient abstract graph spaces, enabling broad searches that would be infeasible to carry out using detailed physical simulations or experiment alone.
The links shown here between graph-theoretic metrics, physical dynamics, and vision performance suggest that similar approaches could be useful across a wider range of network-based physical computing systems.

\begin{figure}[t!]
    \centering
    \includegraphics[width=7in]{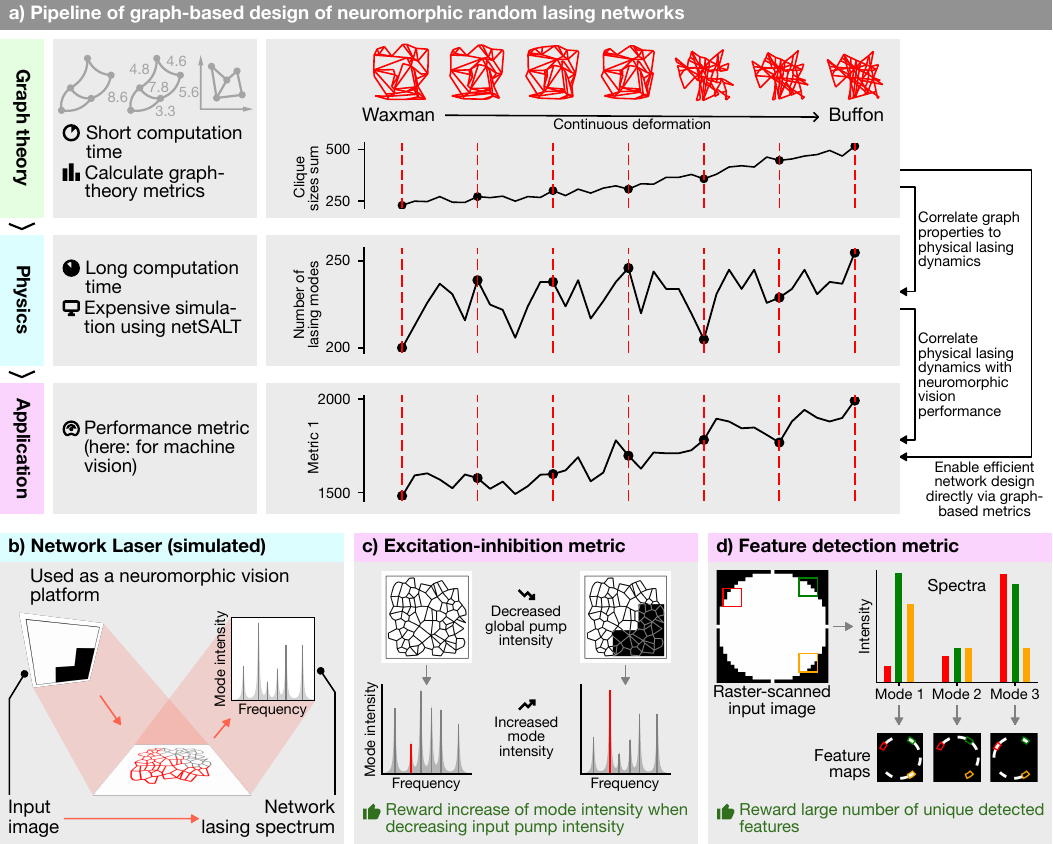}

    \begin{subfigure}{0pt}
        \phantomsubcaption\label{figure1a}
        \phantomsubcaption\label{figure1b}
        \phantomsubcaption\label{figure1c}
        \phantomsubcaption\label{figure1d}
    \end{subfigure}
    \vspace{-\baselineskip}

    \caption{
        \textbf{Working principle of graph-theoretic design of random lasing networks}.
        \textbf{a)} Graph-theoretic analysis pipeline linking graph-based metrics (top) to physical lasing dynamics (middle), and neuromorphic vision performance (bottom). 
        Interpolating between the topologies of two networks, from Waxman (left) to Buffon (right), in graph-metric space reveals associated changes in lasing dynamics (number of lasing modes) and physical vision performance (excitation–inhibition metric), enabling computationally cheap design and optimisation in an abstract graph space that bypasses expensive physical simulations.
        \textbf{b)} Schematic of the random lasing network used as an image processor: input images are projected onto the network via spatially-patterned pump illumination, and the multi-mode lasing spectra recorded as output.
        \textbf{c,d)} Working principles of the vision metrics which quantify the task-level performance of the network. The excitation-inhibition metric (c) measures how many modes increase in intensity when the total illumination power is decreased, arising from strongly-nonlinear lasing mode competition. The feature detection metric (d) quantifies how many distinct feature maps a network can produce, akin to convolutional kernel filters.
    }
    \label{figure1}
\end{figure}

\section*{Results}

The random lasing network which we model here is composed of indium phosphide (InP) waveguides lithographically patterned on a SiO$_2$/Si chip (Figure~\ref{figure1b})~\cite{ng2025fewshotretinomorphicvisionnonlinear,saxena2025designed,dranczewskiPlasmaEtchingFabrication2023a}. 
It serves as a nonlinear physical image processor: projecting an input image onto the network via spatially-patterned pump illumination results in a multi-mode lasing spectrum; feeding the spectrum through a trainable digital output layer (here, a logistic regression layer, see Methods) provides the final computational response (e.g.~which class does an image belong to).
Light generated from spontaneous emission in InP is amplified as it flows through the physical waveguide via optical gain from the pump illumination, and lasing can be reached above a certain gain threshold~\cite{saxenaSensitivitySpectralControl2022,saxena2025designed}.
The lasing processes can be described by steady-state ab initio laser theory (SALT),~\cite{tureciTheorySpatialStructure2007} which is implemented for quantum graphs using the \textit{netSALT}~\cite{saxenaSensitivitySpectralControl2022} framework.
The network hosts a vast set of possible lasing modes, and which subset of these is activated is highly sensitive to the specific image illuminating it.
The output spectrum thus encodes a complex, highly-nonlinear spatially-sensitive representation of the image, allowing the detection of specific image features such as edges, curves or textures\cite{ng2025fewshotretinomorphicvisionnonlinear}, and higher-level processing such as classification or segmentation, with particularly strong performance when training data is scarce\cite{ng2025fewshotretinomorphicvisionnonlinear}.
The specific response and processing imparted by the network is highly dependent on the network topology, which governs the lasing dynamics, spatial mode profiles, and degree of inter-mode coupling.
Designing and optimising this topology to heighten the number of active lasing modes and strength of the coupling is our aim here.

Optimisation of the lasing network topology is performed across three `layers': graph theory, physics, and application (Figure~\ref{figure1a}), with quantitative metrics assessing performance on each.
The underlying abstract mathematical graph comprises nodes and edges, and can include additional structure such as edge weights and spatial coordinates.
We quantify the graph-theoretic parameters of the networks using \textit{highly comparative graph analysis} (HCGA)~\cite{peachHCGAHighlyComparative2021}.
HCGA computes hundreds of graph-theoretic properties for each network design efficiently, allowing \textit{a posteriori} identification of the most relevant parameters by correlating them with metrics describing the lasing dynamics observed in netSALT simulations (including number of lasing modes, strength of inter-mode coupling) and vision performance metrics that we introduce below.

\subsection*{Vision performance metrics}

In order to find correlations between graph-theoretic parameters and physical vision performance without performing expensive machine learning simulations for each candidate network design, we require metrics which assess vision performance while being computationally cheap to evaluate.
Here, we define two \textit{vision/machine learning (ML) performance metrics}.
The `excitation-inhibition' metric (Figure~\ref{figure1c}) counts the number of lasing modes that lase with higher amplitude when the network is partially illuminated, compared against reference spectra where the entire network is maximally illuminated.
It quantifies the excitatory and inhibitory responses in the network, which are key to feature detection processes in biological retina (via lateral inhibition in retinal ganglion cells) and artificial neural networks~\cite{knight_prefrontal_1999,tang_neural_2024,song_training_2016} including lasing networks\cite{ng2025fewshotretinomorphicvisionnonlinear}.
The `feature detection' metric (Figure~\ref{figure1d}) counts the number of distinct edge configurations of a white circle on a black background that can be distinguished from unique lasing spectra output by the network (see Methods).
It is a measure of the number of unique features that the network can detect from images, which is a capability marker of convolutional neural networks (CNNs)~\cite{zhao_review_2024}.

\subsection*{Graph topology predicts nonlinear photonic vision dynamics}
\begin{figure}[h!]
    \centering
    \includegraphics[width=7in]{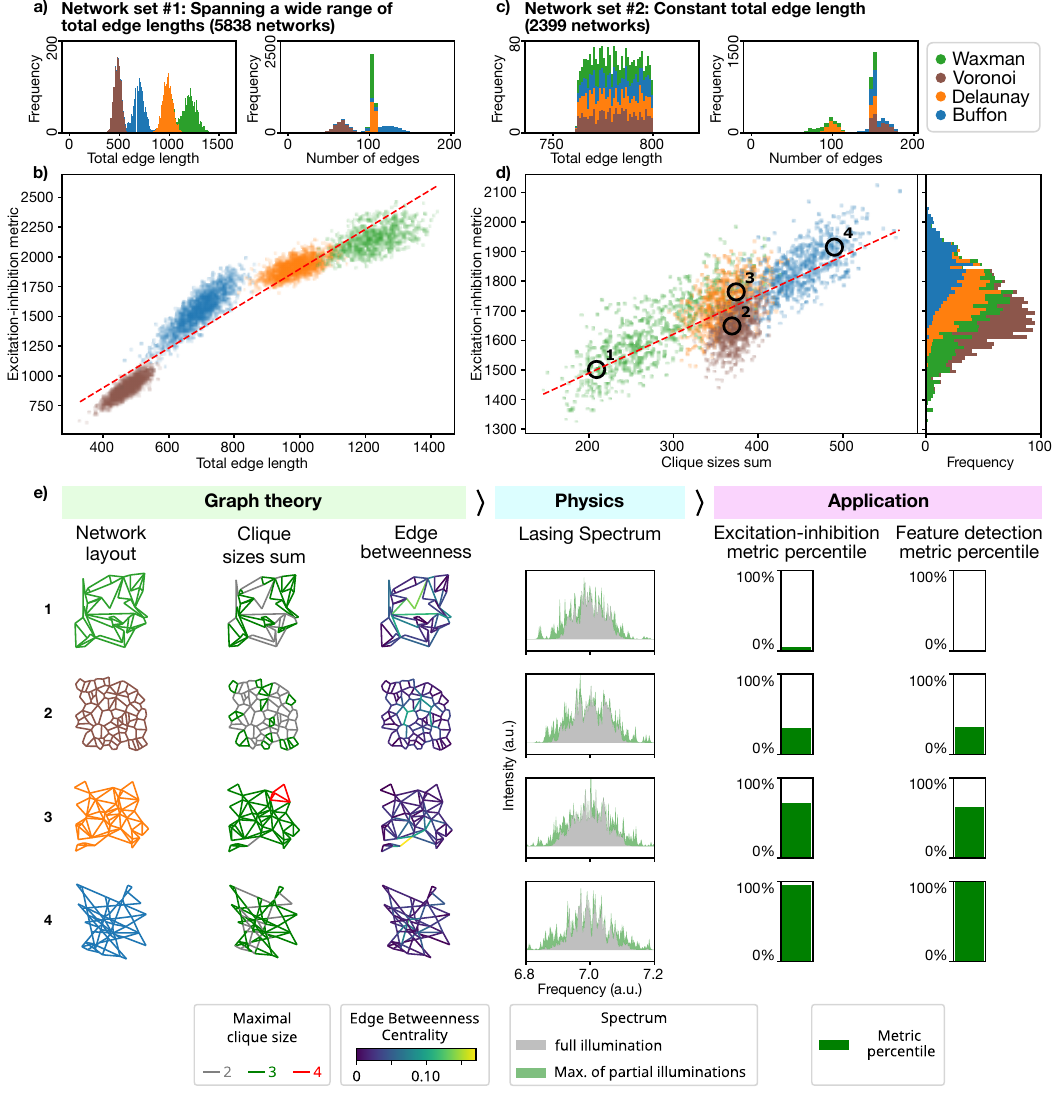}

    \begin{subfigure}{0pt}
        \phantomsubcaption\label{figure2a}
        \phantomsubcaption\label{figure2b}
        \phantomsubcaption\label{figure2c}
        \phantomsubcaption\label{figure2d}
        \phantomsubcaption\label{figure2e}
    \end{subfigure}
    \vspace{-\baselineskip}

    \caption{
        \textbf{Identifying relevant network parameters for physical vision.}
        \textbf{a,b)} In a set of networks (Network set \#1) generated using four different algorithms (Buffon, Delaunay, Waxman and Voronoi) and spanning a wide range of total edge lengths, the excitation-inhibition metric is strongly correlated with the total edge length.
        \textbf{c,d)} Among networks with constrained total edge lengths (Network set \#2), Buffon networks generally perform best on the two vision metrics.
        Among the considered graph-theoretic parameters, \textit{Clique sizes sum} is the best predictor of the excitation-inhibition metric.
        \textbf{e)} Comparative illustration of network topology, graph-theoretic parameters, lasing spectra, and the two vision performance metrics across representative networks generated using the four algorithms (colours correspond to the algorithm legend at the top-right of the figure).
    }
    \label{figure2}
\end{figure}

\begin{figure}[h!]
    \centering
    \includegraphics[width=7in]{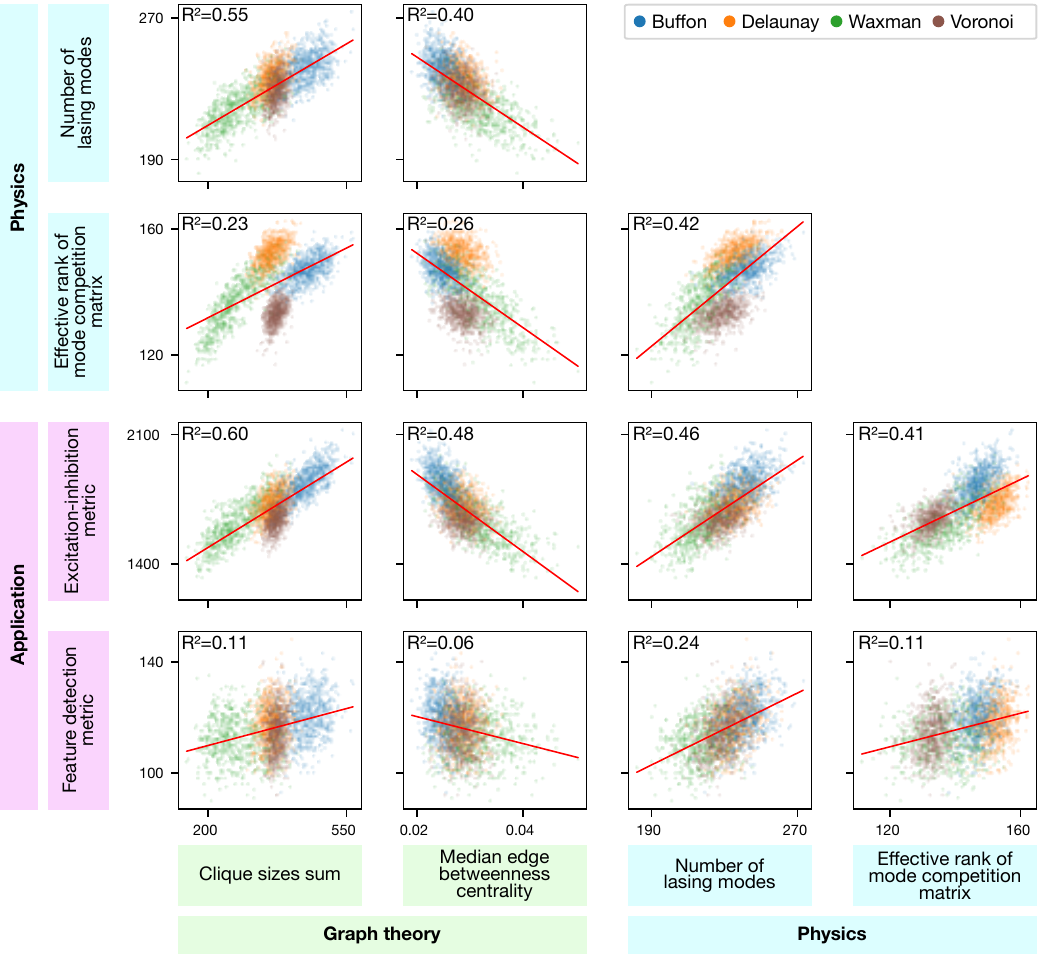}

    \caption{
        \textbf{Relationship between graph-theoretic parameters, physical effects, and machine learning performance}.
        System parameters from graph theory (clique sizes sum and edge betweenness centrality median), physics (number of lasing modes and effective rank of the mode competition matrix), and vision (excitation-inhibition metric and feature detection metric) are mutually correlated in the set of networks with constrained total edge length (Figure~\ref{figure2}). (The same plot for the set of networks with varying total edge lengths can be seen in Supplementary Figure~\ref{figure_s2}.)
    }
    \label{figure3}
\end{figure}

Next, we identify the most relevant graph parameters for physical vision performance by calculating correlations between the graph theory parameters (from HCGA), the physical lasing dynamics (from netSALT), and the two physical vision metrics described above in a diverse set of network topologies.
We construct a large set of random network layouts (Network set \# 1: 5,838 networks) using different generation algorithms: Voronoi\cite{voronoiNouvellesApplicationsParametres1908}, Delaunay\cite{delaunaySphereVideMemoire1934}, Buffon's needle\cite{georges-louisleclercEssaiDArithmetiqueMorale1777}, and Waxman\cite{waxmanRoutingMultipointConnections1988} (Figure~\ref{figure2a}).
All networks are generated subject to geometric constraints to match the limitations of fabrication and experiment (see Methods).
The total edge length of the network $L$ emerges as the parameter having the highest correlation to both vision metrics, with Figure~\ref{figure2b} showing the strong positive correlation with the excitation-inhibition metric (see Supplementary Figure~\ref{figure_s1} for strong positive correlation with the feature detection metric).
Physically, networks with longer total edge length have longer cavities which support more spatially overlapping modes ($\propto L/\lambda$, for a total length $L$ and a light wavelength $\lambda$) within the gain bandwidth,\cite{svelto_passive_2010} enhancing mode competition,\cite{ge_steady-state_2010} which is consistent with the observed correlation.
Maximising the total edge length within the geometrical constraints is thus a simple, physics-informed heuristic to improve vision performance.
In contrast, vision performance is only weakly correlated with the number of nodes in the network (Supplementary Figures~\ref{figure_s3c}-\ref{figure_s3f}).

Most practical physical networks have geometrical constraints including their on-chip footprint, and it becomes infeasible to increase total edge length for a network of fixed diameter beyond a point as waveguides fuse together and heat management become untenable.
Additionally, as the effect of total edge length is strong, keeping it fixed allows us to more readily interrogate the effects of other graph parameters on network dynamics and performance.
We therefore generate a second set of networks (Network set \# 2: 2,399 networks) with a constrained total edge length (Figure~\ref{figure2c}; see Methods for further details).
When total edge length is constrained, the graph parameter `clique sizes sum' (illustrated visually in Fig.~\ref{figure2e}) emerges as the most correlated with the excitation-inhibition metric (Figure~\ref{figure2d}).
Buffon geometry networks exhibit the highest mean clique sizes sum and excitation-inhibition metric, followed by Delaunay and Voronoi which are relatively similar, then Waxman.
Constraining the edge length also reduces the cross-correlation between the two vision metrics, highlighting their independent relevance (Supplementary Figures~\ref{figure_s3a},\ref{figure_s3b}).
The second most correlated graph parameter with the vision metrics is the `median edge betweenness centrality', also illustrated visually in Fig.~\ref{figure2e}.

These two graph parameters which are best correlated with both physical dynamics and vision performance can be understood as follows (see Methods for formal definitions):
\begin{itemize}
  \item \textit{Clique sizes sum:}
        A clique (or complete subgraph) is a set of nodes in which every pair is directly connected by an edge.
        Figure~\ref{figure2e} (second column) visualises the cliques of representative graphs by colouring
          each edge by the size of the largest clique it belongs to; planar graphs cannot host cliques larger than four nodes~\cite{kuratowski_sur_1930}.
        We define the clique sizes sum as the sum of sizes of all cliques of more than two nodes~\cite{peachHCGAHighlyComparative2021,zhangGenomeScaleComputationalApproaches2005}, and find vision performance to be positively correlated with it (Figures~\ref{figure2d},~\ref{figure3}).

  \item \textit{Median edge betweenness centrality:}
        The betweenness centrality
          of an edge is the fraction of node pairs whose shortest path passes through it (Figure~\ref{figure2e}, third column).
        We take the median across all edges~\cite{peachHCGAHighlyComparative2021,brandesVariantsShortestpathBetweenness2008}, and find vision performance to be negatively correlated with it (Figure~\ref{figure3}).
\end{itemize}

Both parameters can be understood intuitively in terms of network connectivity.
A large clique sizes sum reflects a high number of densely interconnected local `neighbourhoods' in the network, with many redundant routes between nearby nodes.
Edge betweenness centrality measures how much of the network's traffic is funnelled through a given edge: a high-betweenness edge is a bottleneck that a large fraction of the network's paths must traverse, whereas a low value indicates that flow is spread across many diverse routes, so that photons and lasing modes explore a richer set of pathways as they propagate.
A network with a large clique sizes sum and a low median edge betweenness is therefore richly connected both \textit{locally} and \textit{globally}, with abundant overlapping paths and few chokepoints.
This is precisely the structure that strengthens the lasing physics: each mode spreads across many waveguide routes at once, and when many modes share the same stretches of the network and spatially overlap, their mode profiles overlap more strongly, driving the mode competition that underpins both vision metrics.
Intriguingly, these measures of connectivity play a major role in other network systems as well.
In urban planning research where street networks are planar graphs, edge betweenness centrality is a standard measure of structural flow that identifies traffic bottlenecks.
Short loops that bypass chokepoints relieve congestion by increasing local path redundancy which is captured by measures closely related to clique sizes sum~\cite{porta2006primal, crucitti2006centrality, kirkley2018betweenness, cardillo2006planar, porta2008parma}.
High-betweenness bottleneck edges are also associated with structural vulnerabilities in social and physical networks~\cite{girvan_community_2002,pournajar_edge_2022}.

The specific physical dynamics which push networks towards better vision performance are highlighted by the strong correlation of the identified graph parameters with metrics of collective network physical dynamics (Figure~\ref{figure3}):
\begin{itemize}
  \item \textit{Lasing mode count} is the number of passive modes of the network that attain lasing threshold under at least one illumination pattern.
      The physics column of Figure~\ref{figure2e} shows the lasing spectra under a full illumination (grey) and the maximum across spectra for all partial illuminations (green).
      \textit{Lasing mode count} is thus the total number of peaks in the graph.
  \item \textit{Mode competition degree} takes the matrix of pairwise overlaps among the modes of the network~\cite{saxenaSensitivitySpectralControl2022} and computes its effective rank\cite{royEffectiveRankMeasure2007} (see Methods).
      Supplementary Figure~\ref{figure_s4} shows examples of mode competition matrices as well as a histogram of \textit{Mode competition degrees}.
\end{itemize}
Both these metrics of lasing physics correlate well with both the identified graph parameters and the vision metrics (Figure~\ref{figure3}).
Well-connected graphs have a larger number of modes within the gain bandwidth with strong spatial overlaps (also see Supplementary Figure~\ref{figure_s5}).
These modes compete strongly for optical gain, increasing the effective rank of the mode competition matrix.
As a consequence, modes which were suppressed under global illumination are able to lase under inhomogeneous spatial illuminations such as specific image inputs, increasing network sensitivity to specific images and hence improving vision processing.
The modes which achieve lasing under inhomogeneous illumination increase total lasing mode count and improve the performance of the network on the excitation-inhibition metric.
Both the number of lasing modes and the effective rank of the mode competition matrix are thus related to the connectivity of the network and affect vision performance.

The `application' column of Figure~\ref{figure2e} shows the metric score on both vision metrics as a percentile score.
The representative Buffon network, which exhibits higher clique sizes sum and lower median edge betweenness than the others, also shows a stronger degree of inhibitive mode coupling and the ability to detect a greater range of distinct image features.
In fact, among the set of graphs with constrained edge length (Network set \#2), Buffon networks have the best average vision performance, while Waxman and Voronoi networks perform the worst (Figure~\ref{figure2d}).
This ordering differs from the unconstrained case (Network set \#1, Figure~\ref{figure2b}) because the unconstrained generators produce networks with very different total edge lengths (Figure~\ref{figure2a}).
The difference in vision performance between generation algorithms can be understood from the behaviour of HCGA graph parameters and lasing dynamics of representative networks (Figure~\ref{figure2e}).
From top to bottom -- Waxman, Voronoi, Delaunay, Buffon -- the clique sizes sum increases and the median edge betweenness centrality decreases, showing an increase in connectivity.
These changes are accompanied by an increase in spectral intensity for partial illuminations and in the number of unique feature maps, showing improved performance according to both vision metrics.

Together, these results illustrate the linked three-layer process depicted in Figure~\ref{figure1a}: graph structure shapes system physics, which in turn determine physical vision performance.

\subsection*{Graph-space evolution optimises networks beyond random design}
\begin{figure}[t]
    \centering
    \includegraphics[width=\textwidth]{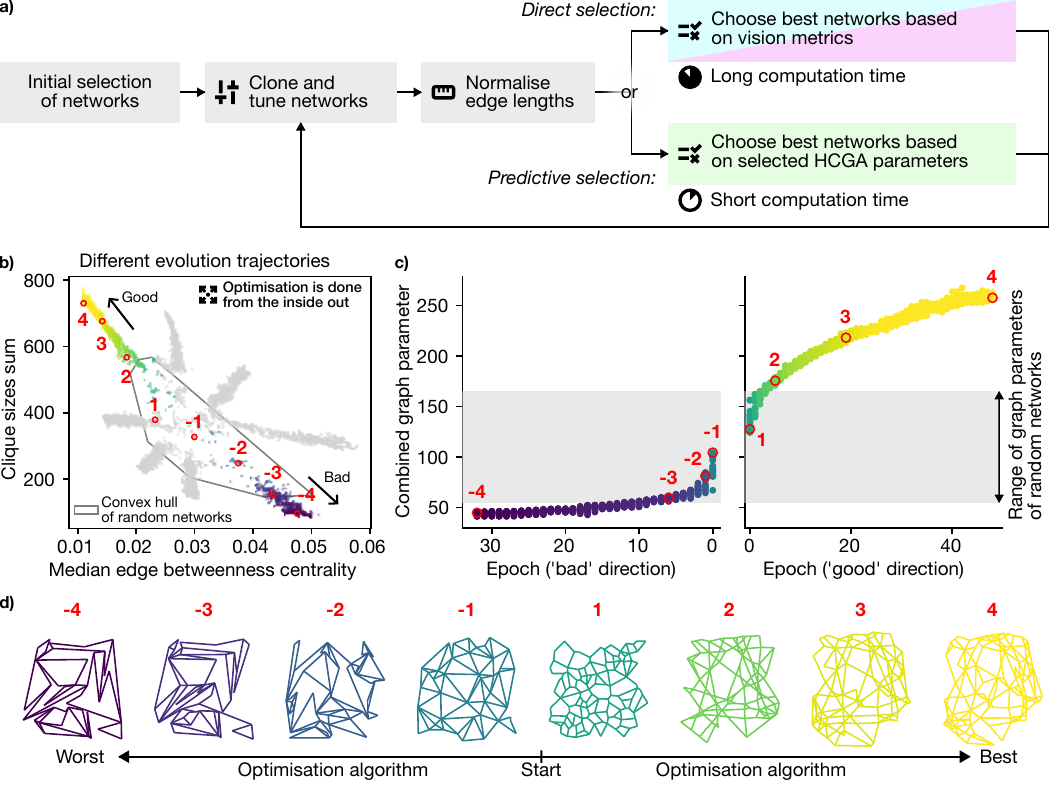}

    \begin{subfigure}{0pt}
        \phantomsubcaption\label{figure4a}
        \phantomsubcaption\label{figure4b}
        \phantomsubcaption\label{figure4c}
        \phantomsubcaption\label{figure4d}
    \end{subfigure}
    \vspace{-\baselineskip}

    \caption{
        \textbf{Evolutionary graph-based optimisation of random lasing networks for physical vision.}
        \textbf{a)} Schematic of the evolutionary algorithm.
        \textbf{b)} Networks can be tuned along arbitrary directions in a two-parameter graph-theoretic space.
        \textbf{c)} The combined graph parameter improves over epochs of the algorithm under predictive selection (for the definition of the combined graph parameter, see Section~\ref{combined_metric}).
        \textbf{d)} Evolution of the network layouts illustrates substantial topological changes.
    }
    \label{figure4}
\end{figure}

Now that we have identified graph metrics correlated to system physics and indicative of vision performance, we may actively use these metrics to optimise networks for physical vision.
We implement an evolutionary algorithm\cite{slowik_evolutionary_2020} that iteratively optimises network topology design over generations by mutating networks and choosing survivors based on a fitness function related to the optimisation target (Figure~\ref{figure4a}, see Methods for details).
Instead of selecting the surviving designs at each generation via expensive netSALT computation of lasing physics and vision metrics (\textit{direct selection}), we perform \textit{predictive selection} which chooses survivors based on the relevant graph parameters (clique sizes sum, and median edge betweenness centrality).
To demonstrate that these metrics are independently tunable, we start with a random selection of networks from Network set \#2 (Figure~\ref{figure2c}) and guide the evolution of network topology along eight cardinal and inter-cardinal directions in a 2D space where $x$ and $y$ axes are the two graph parameters (Figure~\ref{figure4b}).
These directions correspond to increasing, decreasing, or preserving the two parameters independently.
We ensure that the total edge length is fixed throughout the optimisation process (within a 0.5\% tolerance).
Along all explored optimisation directions, the evolutionary algorithm functions well and is able to explore and optimise networks beyond the convex hull of the random starting networks.

To optimise `good' networks for machine vision, we direct network topology towards the direction in graph parameter space that improves both vision metrics: increasing clique sizes sum, and decreasing median edge betweenness centrality.
We perform this evolutionary optimisation using the predictive selection scheme which performs no netSALT simulations, and only evaluates the computationally cheap graph parameters.
Under evolutionary optimisation in this `good' direction for 49 epochs (Figure~\ref{figure4c}), the fitness function combining both graph parameters (see Section~\ref{combined_metric}) clearly exceeds the range achieved by the starting set of 100 random networks (shaded in grey).
We also perform optimisation in the opposite `bad' direction and show the corresponding decrease in combined graph parameters over 33 epochs.
We select eight networks from these processes, four increasingly `good' networks (labelled 1 to 4) and four increasingly `bad' (labelled -1 to -4), showing visible changes in topology and node connectivity along the direction of evolutionary design (Figure~\ref{figure4d}).

We perform diagnostic evaluation of the vision metrics for a fraction of the optimised networks and confirm that the outlined process works: optimised networks with improved graph parameters show corresponding improvements in vision metrics (Supplementary Figure~\ref{figure_s7a}).
In fact, they achieve higher vision metric values than those of all networks in Network set \#2 (Figure~\ref{figure2c}).
This is a substantial result, and clearly demonstrates the efficacy of the directed evolution approach using predictive selection.
Predictive selection is also substantially faster than direct selection (Figure~\ref{figure4a}) as it does not require netSALT evaluation. Tuning one network and evaluating its vision metrics using netSALT requires an average of 43,000 CPU-core seconds, whereas tuning one network and computing its graph parameters requires 14.5 CPU-core seconds. Predictive selection therefore reduces the per-candidate evaluation time by approximately 3000× relative to netSALT-based selection. All timings were normalised to a single CPU core; the calculations themselves were distributed using parallel computing.

\subsection*{Graph-optimised networks reduce classification error on vision tasks}

\begin{figure}[h!]
    \centering
    \includegraphics[width=0.9\textwidth]{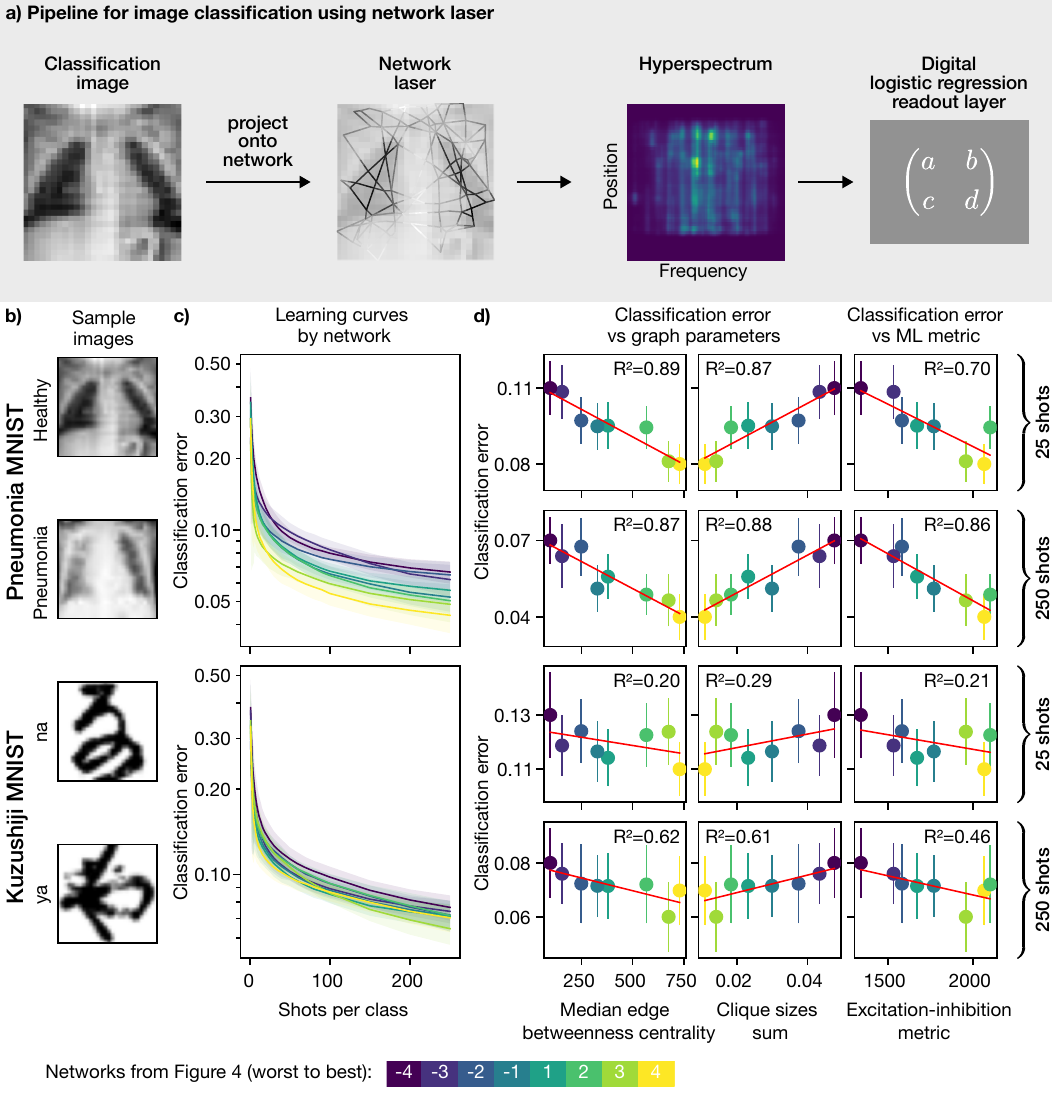}

    \begin{subfigure}{0pt}
        \phantomsubcaption\label{figure5a}
        \phantomsubcaption\label{figure5b}
        \phantomsubcaption\label{figure5c}
        \phantomsubcaption\label{figure5d}
    \end{subfigure}
    \vspace{-\baselineskip}

    \caption{
        \textbf{Demonstrating improved performance in machine vision tasks.}
        \textbf{a)} Schematic of the pipeline for image classification using a lasing network.
        \textbf{b)} Sample images from both classes of both classification tasks.
        \textbf{c)} Learning curves of networks from the evolutionary optimisation algorithm. The ML error tends to decrease along the optimisation direction.
        \textbf{d)} Relationship between the classification error and both graph parameters as well as one vision metric for two different number of shots per class.
    }
    \label{figure5}
\end{figure}

We now evaluate the performance of the optimised networks on two separate two-class image classification tasks in-silico: Pneumonia MNIST\cite{medmnistv1,medmnistv2} and Kuzushĳi MNIST\cite{clanuwat2018deep} (Figure~\ref{figure5b} and Supplementary Figure~\ref{figure_s6}).
Images from the datasets are projected onto the networks in netSALT simulations, generating output hyperspectra (see Methods) that are then fed into a logistic regression classifier layer to predict the class (Figure~\ref{figure5a}).
As ML pipelines require large numbers of images to be processed, this is a computationally expensive step due to the corresponding increase in netSALT simulations.
We therefore evaluate a sampled subset of networks, namely the eight networks shown in Figure~\ref{figure4d}.
We compute the ML classification error (1 - accuracy) for different numbers of shots per class $S$ used for training, ranging between 1 and 250 (see Methods).

Classification error generally decreases for the networks sampled towards the `good' optimisation direction for both vision tasks (Figure~\ref{figure5c}, colours correspond to the selected networks in Fig.~\ref{figure4d}).
Correlating the classification error against the graph parameters and vision metrics for two selected shot sizes illustrates the improvement more clearly (Figure~\ref{figure5d}).
For the Pneumonia MNIST task, we observe strong correlation between classification error and the graph parameters, with $R^2$ values of 0.88-0.89.
This in fact slightly exceeds the correlation between the classification error and the excitation-inhibition metric ($R^2 = 0.7-0.86$), further justifying the use of predictive selection for optimisation.
Error bars are drawn from multiple repeated classification tests, with different shuffles of images into training and test sets (see Methods).

For the Kuzushiji MNIST task, the correlations between metrics and classification error are lower for $S=25$, with 
$R^2 = 0.2-0.29$ for graph metrics and $R^2=0.21$ for the excitation-inhibition metric.
The diversity of images within each class makes low-data classification challenging and obscures correlation, with high variability dependent on the specific selection of images in the training set.
The case of $S=250$ exhibits substantially higher correlation as the larger training set allows for more stable training, with the graph metrics showing good correlation ($R^2 = 0.61-0.62$), again higher than the vision metric ($R^2=0.46$).
These results confirm the correspondence between the predictive graph metrics and ML performance on image classification tasks, with the best evolutionary optimised network design (network 4, yellow) outperforming all other network designs in all cases on the Pneumonia MNIST task, and performing best in the 25 shot Kuzushiji MNIST task and second best in the 250 shot case (network 3 performs best here).

The results in Fig.~\ref{figure5} support the choice of the heuristic excitation-inhibition vision metric as they do exhibit correlation with machine vision performance, albeit weaker than the graph parameters. Developing vision metrics that include the spatial lasing information to better match the hyperspectral experiment and identifying associated graph-theoretical parameters may further improve correlations.

Our results demonstrate that the evolutionary algorithm (Figure~\ref{figure4a}) can efficiently improve network topology design and vision performance purely via graph-theoretical parameters.
Fabrication and characterisation of the optimised networks is the next step to validate the enhanced performance relative to the randomly selected networks~\cite{ng2025fewshotretinomorphicvisionnonlinear}.

\section*{Conclusions}

We have demonstrated a graph-theoretic procedure for the optimisation of network-based physical systems, and demonstrated its utility to design random lasing networks for neuromorphic vision.
In particular, we have demonstrated the efficient search and correlation of graph parameters against a range of physical dynamics and physical vision performance, and developed an evolutionary algorithm that uses graph-based fitness to efficiently optimise network design and topology for objectives which are computationally expensive to evaluate directly.

We have shown that nonlinear lasing dynamics in random lasing networks -- including mode-coupling, inhibitive inter-mode competition, and the number of lasing modes -- are strongly correlated with abstract graph-based metrics.
This relationship provides a route towards understanding the complex emergent dynamics of these systems and motivates more detailed physics-based studies. We have additionally established correlations between graph parameters and functional vision performance, enabling the rapid and computationally efficient design of candidate network topologies for experimental fabrication and validation.
The best networks for physical vision have two features: large total edge lengths to support more modes within the gain bandwidth, and a high degree of connectivity to maximise the spatial overlap and nonlinear interactions between the modes.
By running a set of optimised network topologies through an ML image classification pipeline across two tasks, we demonstrated the ability of our evolutionary algorithm to enhance ML classification performance solely via abstract graph parameters at radically higher computational efficiency than detailed physical simulations.

Our results broaden the suite of efficient tools for understanding and optimising the network topology and physical dynamics of strongly-interacting physical networks, with specific contributions to the design of networks for physics-based neural networks and computational substrates.
We anticipate that the graph-based approach here will open routes towards promising and optimised platforms for physics-based machine learning, and an enhanced understanding of the nuanced links between graph theory, network design, and emergent nonlinear dynamics in network-based physical systems.
We believe that a vast range of other technologically relevant network-based systems may benefit from similar graph-based approaches, spanning diverse fields such as biochemistry\cite{patel_graph_2024}, energy transmission\cite{emery_complex_2024}, and materials~\cite{lawrie_application_2024}.

\subsection*{Author contributions}

Conceptualisation: TVR, RS, JCG, TF, JD, WKN, MB;
Methodology: PO, TVR, JCG, MB, JD, WKN, AA, RLP, TF;
Implementation: PO, YY, TS, EG;
Investigation: PO, YY, JD, TF, WKN, TVR, JCG, RS;
Visualisation: PO;
Funding acquisition: RS, JCG;
Supervision: TVR, JCG, RS, TF, JD, WKN;
Writing: PO, JCG, YY, TVR with contributions from all authors.

\subsection*{Acknowledgements}

We acknowledge computational resources and support provided by the Imperial College Research Computing Service (10.14469/hpc/2232).
TVR and RS acknowledge support from the Engineering and Physical Sciences Research Council (EPSRC), grant number EP/T027258 and EP/Y015673.
JD acknowledges support from the EU ITN EID project CORAL (GA no. 859841).
JCG, RS and JD acknowledge support from the Imperial College London President's Excellence Fund for Frontier Research.

\subsection*{Competing interests}

The authors declare no competing interests.

\subsection*{Data availability statement}

The datasets generated during and/or analysed during the current study are available from the corresponding author on reasonable request.

\subsection*{Code availability statement}

The code used in this study is available from the corresponding author on reasonable request.

\section*{Methods}

\subsection*{Random lasing network as image preprocessor}

An input image is projected onto the lasing network using a digital micro mirror device and induces spatially inhomogeneous gain that results in lasing~\cite{ng2025fewshotretinomorphicvisionnonlinear}.
The output is a hyperspectrum consisting of spatially resolved lasing spectra, with one wavelength dimension and one spatial dimension.
To solve machine learning problems with this setup, one has to feed the hyperspectrum as the input of a small, final trainable linear layer.
In the experimental setup, this last layer is implemented in-silico.
Thus, the physical lasing network system serves as an image preprocessor whose rich nonlinear dynamics is exploited by a purely linear layer in the end.
Although the experimental setup~\cite{ng2025fewshotretinomorphicvisionnonlinear} uses the spatially and spectrally resolved hyperspectrum, we consider only the intensities of the lasing modes in this manuscript for simplicity, until the final image classification step.

\subsection*{Network generation and constraints}

The networks have physical and manufacturing constraints -- besides the networks being planar, edges have constraints on minimum length ($1.5\ \si{\micro m}$) and minimum angle ($5\si{\degree}$); nodes have constraints on minimum separation from edges ($0.5\ \si{\micro m}$) and maximum degree ($7$) to preserve the node shape with finite-width waveguides.
Additionally, the networks have to fit in a square of side length $S=50\ \si{\micro m}$ for reasons of comparability and limitations of computational resources.
These constraints are always enforced during generation; a given generation algorithm is iterated until it produces a constraint-satisfying network.

We generate two sets of networks (Figure~\ref{figure2a},\ref{figure2c}) and use four different generation algorithms (Table~\ref{tab:network-generation-algorithms}). In the case of the second set of networks, the listed numerical values were experimentally chosen to create networks within the desired range of total edge length.  We limit ourselves to closed networks -- those without dangling unconnected waveguides/edges with one end not connecting to a node -- to reduce the effects of energy dissipation.
\begin{table}[h!]
    \caption{Description of network generation algorithms. When parameters are indicated to be chosen randomly, they are chosen for each generated network individually.}
    \label{tab:network-generation-algorithms}
    \centering{
        \begin{tabular}{p{1.5cm}p{7.5cm}p{2.2cm}p{4.6cm}}
            \toprule
            \multirow{2}{*}{\textbf{Name}} & \multirow{2}{*}{\textbf{Description}} & \multicolumn{2}{l}{\textbf{Parameters}} \\
             &  & \begin{tabular}[c]{@{}l@{}}Network set \#1\end{tabular} & \begin{tabular}[c]{@{}l@{}}Network set \#2\end{tabular} \\ \midrule
            Voronoi & Randomly place $N$ points within the allowed square. Generate the Voronoi graph corresponding to these points, and remove all edges that do not lie completely within the bounding square. & $N=60$ & $N \sim \text{Unif}(\{25,\dots,140\})$ \\ \midrule
            Delaunay & Randomly place $N$ nodes within the allowed square one-by-one ensuring that the minimum edge length is always satisfied. Generate the Delaunay graph based on the set of nodes. & $N=40$ & $N \sim \text{Unif}(\{40,\dots,55\})$ \\ \midrule
            Buffon's needle & Randomly place $N$ infinitely long needles by uniformly choosing a point on the needle within the square and an angle in $[0,\pi]$. Create the Buffon graph created by these needles and remove all edges that do not lie completely within the square. & $N=20$ & $N \sim \text{Unif}(\{23,\dots,27\})$ \\ \midrule
            Waxman & Randomly place $N$ nodes within the allowed square in the same way as for Delaunay. Iteratively add $E$ edges between pairs of nodes: at every step, consider all potential constraint-satisfying edges, weigh the random selection of each potential edge by $e^{-l/L}$ where $l$ is the length of the potential edge. Start over if no valid edges remain. \textit{Note that this algorithm is similar to the Waxman algorithm\cite{waxmanRoutingMultipointConnections1988}, but not identical due to the fixed number of edges and the existence of network constraints.} & $L=\infty$, \newline $N=50$, \newline $E=105$ & $L \sim \text{Unif}([1,16])$, \newline $N \sim \text{Unif}(\{50,\dots,60\})$, \newline $E \sim 115 - \text{Unif}(\{\lfloor3L\rfloor, \lceil2L\rceil\})$ \\ \bottomrule
        \end{tabular}
    }%
\end{table}

\subsection*{netSALT}

The lasing response of the networks is modelled using a modified version of netSALT (details of the theory are available in Ref.~\citeonline{saxenaSensitivitySpectralControl2022}).
The designed networks are represented as graphs, with the edges subdivided to overlap with a $3\times3$ grid to correspond to the illumination pattern.
Edges have a complex refractive index n = 3.4 + 0.001i when unpumped, have a Lorentzian gain profile centred at frequency $k_a = 7$ with width $\gamma_a = 0.35$, and are pumped up to a power $D_0 = 0.5$ (For a detailed physical explanation of these normalised parameters, see Ref.~\citeonline{saxenaSensitivitySpectralControl2022}).
The passive (unpumped) modes of the network are first identified in the complex frequency plane $k$.
A separate simulation is then performed for each $3\times3$ binary pattern by pumping only the illuminated edges, and the complex frequency of every mode from the passive simulation is tracked as a function  of pump power.
A mode attains threshold and starts lasing when its frequency becomes real; beyond this pump power, its frequency and shape do not change -- but all the lasing modes compete for gain to modify their lasing intensities and further raise the thresholds of remaining modes.

We sped up netSALT in three ways to allow for large-scale simulations:
i) As tracking modes as a function of pump power is computationally expensive, we identify modes early with a gain variation that is too low to attain lasing threshold and neglect them.
ii) Instead of converging to the exact lasing threshold of modes through a procedure with adaptive step sizes, we use a fixed optimal step size, allow to overshoot the threshold and perform linear interpolation afterwards.
iii) We optimised parallelism.
All these improvements were validated for accuracy on individual simulations by comparing with the original code before performing the simulations in the manuscript.

As netSALT does not provide linewidths of the lasing modes, the realistic spectra in Figure~\ref{figure2e} are created by assigning a Lorentzian width of 0.005 to each mode.
The feature maps in Figure~\ref{figure2e} are made by selecting specific modes and plotting their intensities for the patterns corresponding to the kernel scan of the image of a circle.
The simulated hyperspectra in Figures~\ref{figure1b} and~\ref{figure1c} approximate the bulk of the experimental hyperspectra as arising from network nodes.
We assign each node a fixed radius and compute a spatial distribution of energy dissipation linked to each mode from its intensity profile across nodes. 
Summing over all modes and assuming a finite linewidth, we get a hyperspectrum of dimension 3 that assigns each tuple $(x,y,f)$ an intensity ($x$ and $y$ are spatial coordinates, and $f$ is the frequency coordinate). Now, by numerically integrating over the $y$ dimension, we finally get a hyperspectrum of dimension 2 that assigns each tuple $(x,f)$ an intensity.

\subsection*{Vision metrics}

\subsubsection*{Excitation-inhibition metric}

We start with a $128 \times 128$ pixel image of a white circle on black background.
From this, we extract the 45 unique $3 \times 3$ windows where at least three pixels are white (we assume the lasing spectrum to be completely dark for the rest).
We illuminate every pattern $\mathcal{P}$ onto a network $\mathcal{N}$ using netSALT, which outputs for each lasing mode $\mathcal{M}$ and for each pattern, the lasing intensity $I(\mathcal{N}, \mathcal{P}, \mathcal{M})$.
For each partial illumination (at least one dark pixel), we count the number of modes that are stronger than for a full illumination. We define the excitation-inhibition metric as the sum of these counts across all partial illuminations,
\begin{equation}
    m_1(\mathcal{N}) := \sum_{\mathcal{P}} \# \left\{ \mathcal{M} : I(\mathcal{N}, \mathcal{P}, \mathcal{M}) > I(\mathcal{N}, \mathcal{P}_{full}, \mathcal{M}) \right\}.
\end{equation}

\subsubsection*{Feature detection metric}

We find convolutional feature maps of the circle for each lasing mode using the spectra computed above.
We consider only feature maps that perform edge detection by having a higher intensity at the edge than in the interior of the circle.
The remaining feature maps look schematically like in Figure~\ref{figure1d}, where each mode detects edges only along certain directions.
We describe the oriented edge detection capability of a lasing mode in a network by a 16-element boolean vector $f(\mathcal{N}, \mathcal{M})$.
The $k$th element of the vector is 1 if the feature map is above a certain threshold in the angle range $\left[\frac{k}{16} \cdot 2\pi, \frac{k+1}{16} \cdot 2\pi\right)$.
Examples of $f(\mathcal{N}, \mathcal{M})$ are shown in Figure~\ref{figure2e}.
Let $S(\mathcal{N})$ be the set of unique vectors among $f(\mathcal{N}, \mathcal{M})$.
The feature detection metric is now defined by:
\begin{equation}
    m_2(\mathcal{N}) := \left[ \# S(\mathcal{N}) \middle] \cdot \middle[ \frac{\dim{(\text{span }{S(\mathcal{N})})}}{16} \middle] \cdot \middle[\frac{\sum_k{\max_\mathcal{M}{f_k(\mathcal{N}, \mathcal{M})}}}{16} \right].
\end{equation}
The first term in brackets rewards distinct feature vectors, the second one rewards their linear independence, and the third one rewards edge detection along all directions.

\subsection*{Network parametrisation and statistical analysis}

We consider the networks as weighted graphs with the physical length of the edge as its weight.
We reduce the graphs by iteratively removing vertices of degree two and combining their incident edges, as the resulting graphs are equivalent.
There is one case where this reduction is impossible: a triangle with nodes $u$, $v$, and $w$, where $w$ has degree 2; one cannot unify the edges $\{u,w\}$ and $\{v,w\}$ as the nodes $u$ and $v$ are already connected.
We retain such edges as HCGA does not allow for multigraphs.
We parametrise the reduced networks using HCGA, while the physical simulations using netSALT are performed using the original network since the embedding into real space is physically relevant.

After computing the HCGA graph parameters and performance metrics for all networks, we calculated the Pearson correlation coefficients\cite{wassermanAllStatisticsConcise2004} between them and identified the graph parameters correlated the strongest with the metrics.
Usually, a graph parameter is correlated similarly to both metrics as both metrics are correlated to each other quite strongly (see Supplementary Figure~\ref{figure_s3c} to~\ref{figure_s3f}).

As presented above, we identified two graph parameters which are strongly correlated with both the vision metrics and the metrics of physical lasing dynamics: \textit{clique sizes sum} and \textit{median edge betweenness centrality}. In the following, we will define these rigorously.

\subsubsection*{Clique sizes sum}

Let the network at hand be given by a connected undirected graph $G=(V,E)$, with vertex set $V$ and edge set $E \subseteq \binom{V}{2} := \{e\subseteq V : |e|=2\}$. We call $V' \subseteq V$ a \textit{clique} of $G$ if $\binom{V'}{2} \subseteq E$. In words: A clique of $G$ is a subset of vertices that are connected pairwise in the whole graph. Denote the set of all cliques of $G$ by $C(G)$. The \textit{clique sizes sum} is defined by
\begin{equation}
    \sum_{V' \in C(G)}{\mathbbm{1}_{|V'| > 2} \cdot |V'|}.
\end{equation}

\subsubsection*{Median edge betweenness centrality}

For two distinct nodes $v_1, v_2 \in V$ and an edge $e \in E$, let $\sigma(v_1,v_2)$ be the number of shortest path between $v_1$ and $v_2$ (in terms of number of edges) and let $\sigma(v_1,v_2 | e)$ be the number of those paths that go through $e$. The (normalised) \textit{edge betweenness centrality} of the edge $e$ is defined by
\begin{equation}
    \text{EBC}(e) := \frac{1}{|V|(|V| - 1)} \sum_{\substack{v_1,v_2 \in V \\ v_1 \neq v_2}} {\frac{\sigma(v_1,v_2 | e)}{\sigma(v_1,v_2)}}.
\end{equation}
Roughly speaking, $\text{EBC}(e)$ counts what proportion of shortest paths within $G$ travel through the edge $e$.

\subsection*{The mode competition matrix and its effective rank}

Mode competition $M$ is a measure of the spatial overlap between the different modes for a specific illumination. It is defined as\cite{saxenaSensitivitySpectralControl2022}
\begin{equation}
    M_{\mu\nu} := \Gamma_\nu \text{Real}\left( \frac{\int{|u_\nu(x)|^2u_\mu(x)^2 \delta_{\text{pump}}(x) \text{d}{x}}}{\int{u_\mu(x)^2 \delta_{\text{pump}}(x)\text{d}x}} \right),
\end{equation}
where $\mu,\nu$ are mode indices, the integrals are over all edges, $\delta_{\text{pump}}(x)$ denotes the pump profile, $u_\mu(x)$ is the mode profile (i.e. the normalised electric field along the edges), and $\Gamma_\mu$ is the gain linewidth.

Now, let $M$ be the mode competition matrix for a full illumination and let $M'$ be this matrix with diagonal entries set to zero. Both $M$ and $M'$ are symmetric; hence, they can be diagonalised with real eigenvalues. Let the eigenvalues of $M'$ be denoted by $\lambda_i$. The effective rank of $M'$ is defined as\cite{royEffectiveRankMeasure2007}
\begin{equation}
    \text{erank}(M'):=\exp(-\sum_i{p_i \log{p_i}}) \text{ where } p_i := \frac{|\lambda_i|}{\sum_i{|\lambda_i|}}.
\end{equation}
When we refer to the effective rank of the mode competition matrix, we mean $\text{erank}(M')$.

Supplementary Figure~\ref{figure_s4} shows examples of mode competition matrices as well as a histogram of their effective ranks.

\subsection*{Network tuning and renormalisation}

\subsubsection*{Network mutation}

Network mutation is done in steps; at each step, a different kind of tuning operation is chosen at random. The tuning operations are: adding a node and some edges to existing nodes; removing a node and all adjacent edges; adding an edge between existing nodes; removing an edge. All tuning operations respect network constraints. After a bit of experimentation, we gave the tuning operations statistical weights of $1.00$, $0.25$, $1.00$, and $2.00$, respectively.

\subsubsection*{Network renormalisation}

After performing a tuning operation, we \textit{renormalise} the network to keep the total edge length same as before.
This is done using gradient descent by expressing the total edge length as a function of the $x$- and $y$-coordinates of all nodes.
The network is iteratively returned to within 0.5\% of its initial total edge length by making a maximum change of 0.2\% at each step.
If some network constraint (except planarity) is violated after one gradient descent step, the compliance is iteratively restored by moving as few nodes as possible.
For example, if an edge becomes too short, both of its end nodes are moved slightly apart.
If the network ceases to be planar after some iteration step, the renormalisation process is reset and restarted with a smaller step size.

\subsection*{Network interpolation and optimisation}

\subsubsection*{Network interpolation}

When interpolating between two networks A and B, treat them as members of an $n$-dimensional space spanned by HCGA parameters. In this paper, we use $n=2$ and the HCGA parameters \textit{clique sizes sum} and \textit{median edge betweenness centrality}.
The HCGA parameters are linearly normalised such that they have standard deviation $1$ in the second set of networks (Figure~\ref{figure2c}) to compute Euclidean distances in HCGA space.
We perform bidirectional interpolation over a pool of networks (initially just A and B) epoch-by-epoch; with each epoch involving:\cite{slowik_evolutionary_2020}
\begin{enumerate}
    \item Amend the pool by tuning and renormalising each network. Label each tuned network with A or B, depending on which network it originated from.
    \item Iteratively choose the $N$ A-B labelled pairs of networks that are closest in $n$-dimensional HCGA space. Delete all but these $2N$ networks.
\end{enumerate}
The process terminates when the pairwise distances fall below 1\% of initial distance. We used $N=10$.

\subsubsection*{Network optimisation into cardinal directions}

We optimise networks into different cardinal directions in HCGA space to demonstrate that the chosen graph parameters are tunable independently up to a certain degree (see Figure~\ref{figure4b}).
Network optimisation is performed similar to network interpolation, using an evolutionary algorithm\cite{slowik_evolutionary_2020}.
We start with a set of $N$ networks labelled $G_1, \dots,G_N$ that are randomly chosen from the set of random networks of constant edge length, and treat them in $n$-dimensional HCGA space $\mathbb{R}^n$.
Let $\vec{x}_0 \in \mathbb{R}^n$ denote the average position of the initial networks, and $\vec{v} \in \mathbb{R}^n$ with $||\vec{v}||=1$ be the desired cardinal direction of optimisation (see Figure~\ref{figure4b}).
The optimisation is done epoch by epoch; with each epoch involving:
\begin{enumerate}
    \item For each network $\mathcal{N}$, calculate its position $\vec{x} \in \mathbb{R}^n$ in HCGA space, and then assign a score $S$ that rewards motion in the desired direction and penalises other directions:
    \begin{equation}
        S(\mathcal{N}) := \Delta\vec{x} \cdot \vec{v} - \big|\big|\Delta\vec{x} - (\Delta\vec{x} \cdot \vec{v})\vec{v}\big|\big| \text{ where } \Delta\vec{x} := \vec{x}-\vec{x}_0.
    \end{equation}
    \item Iteratively choose $M$ networks with the highest score.
    After choosing a network, apply a penalty to the scores of all other networks with the same label to reward diversity.
    For a given label, the penalties are cumulative but the additional penalty value halves at each application.
    Delete all but the $M$ chosen networks.
    \item Amend the pool of networks by tuning and renormalising each network $\frac{N}{M}-1$ times, yielding $N$ networks in total. Label each tuned network with $G_1$,~$G_2$,~$\dots$,~or~$G_N$ depending on which network it originated from.
\end{enumerate}
We used $n=2$, $N=100$ and $M=20$. The cardinal directions $\vec{v}$ of optimisation were chosen based on the spread of clique sizes sum and median edge betweenness centrality in the random set of networks of constant edge length. We ran the optimisation algorithm for 48 epochs into all directions except for \textit{decreasing} clique sizes sum and \textit{increasing} median edge betweenness centrality where tuning steps stopped to converge after 32 epochs already.

\subsubsection*{Network optimisation into `good' direction}

Based on our previous findings, we want to \textit{increase} the networks' clique sizes sum and \textit{decrease} their median edge betweenness centrality in order to increase their vision metrics. Of course, this could also be done by defining a cardinal direction $\vec{v} \in \mathbb{R}^2$ in HCGA space accordingly and then applying the method above. However, for such a multi-objective optimisation problem, it is more natural to make use of the Pareto front\cite{colletteMultiobjectiveOptimizationPrinciples2004}: A network is defined to be on the Pareto front of a given set of networks if no other network within this set has both a higher clique sizes sum and a lower median edge betweenness centrality. The 2nd order Pareto front of a set of networks can be obtained by first removing the first Pareto front from the set and then computing the Pareto front of the remaining set. This way, one can iteratively define the $n$-th order Pareto front of a set of networks. For each network in a set, we can determine the unique order of the Pareto front that it is a part of. The lower this order, the better the network.

Our network optimisation into the `good' direction works just like the network optimisation into cardinal directions with the only difference that the score is defined like so:
\begin{equation}
    S(\mathcal{N}) := -(\text{unique order of the Pareto front that $\mathcal{N}$ is part of}).
\end{equation}

\subsection*{ML pipeline for image classification}

For each of the two datasets, Pneumonia MNIST\cite{medmnistv1,medmnistv2} and Kuzushĳi MNIST\cite{clanuwat2018deep}, we projected 800 images (400 per class) of size $28 \times 28$ onto each network using netSALT (see Figure~\ref{figure5a}), yielding hyperspectra.
The hyperspectra are rasterised as follows: A single global frequency crop window is determined by summing the frequency spectra of all hyperspectra (for each network and each dataset) and selecting the smallest frequency interval that contains 99.5\% of the total spectral intensity. Every hyperspectrum is subsequently cropped to this identical frequency range to ensure consistent frequency alignment across all samples. The cropped hyperspectra are then downsampled by dividing the frequency axis into 128 bins and replacing each bin with its average intensity. Similarly, the spatial axis is averaged into 64 equally sized bins.

After rasterisation, the hyperspectra were flattened into feature vectors, and used as inputs to a logistic regression classifier.
Since the full 8192-dimensional feature space contains many redundant or noisy features, we first perform feature selection, retaining the top $k$ features.
The number of selected features, $k$, together with the inverse regularisation strength, $C$, of the logistic regression model, are optimised using Optuna~\cite{akiba2019optuna,scikit-learn}.
To reduce computational cost, this ($k$, $C$) optimisation is performed only once at 200 shots  (see below) for each dataset and network, and the resulting optimal hyperparameters ($k$, $C$) are reused for all other shot numbers.

We train the network with different shot sizes $S$ ranging from 1 to 250.
For each shot size, the training procedure is repeated in $T$ trials to suppress the effects of statistical fluctuations, with more trials for smaller shot sizes (see Table~\ref{tab:repeats}).
In each trial, $S$ randomly selected images per class form the training set, while the remaining $N = 800 - 2S$ images form the test set.
The classifier is trained and evaluated on each trial, and the classification error for the shot size (Figure~\ref{figure5c}) is taken to be the average over all trials. The error bars in Figure~\ref{figure5d}) are the standard deviation of the classification errors between all trials.

\begin{table}[h]
\centering
\begin{tabular}{cc}
\toprule
Shots per class ($S$) & number of trials ($T$) \\
\midrule
1--7   & 750 \\
8--11  & 600 \\
12--99 & 375 \\
$\geq100$ & 225 \\
\bottomrule
\end{tabular}
\caption{Number of trials used for each shot count.}
\label{tab:repeats}
\end{table}

\nocite{blondelFastUnfoldingCommunities2008}

\bibliography{references}

\setcounter{figure}{0}
\setcounter{section}{3}
\renewcommand{\thefigure}{S\arabic{figure}}
\renewcommand{\thesection}{S\arabic{section}}

\refstepcounter{section}
\label{combined_metric}

\refstepcounter{figure}
\label{figure_s1}
\refstepcounter{subfigure}
\label{figure_s1a}
\refstepcounter{subfigure}
\label{figure_s1b}
\refstepcounter{subfigure}
\label{figure_s1c}
\refstepcounter{subfigure}
\label{figure_s1d}
\refstepcounter{subfigure}
\label{figure_s1e}
\refstepcounter{subfigure}
\label{figure_s1f}

\refstepcounter{figure}
\label{figure_s2}
\refstepcounter{subfigure}
\label{figure_s2a}
\refstepcounter{subfigure}
\label{figure_s2b}
\refstepcounter{subfigure}
\label{figure_s2c}
\refstepcounter{subfigure}
\label{figure_s2d}
\refstepcounter{subfigure}
\label{figure_s2e}
\refstepcounter{subfigure}
\label{figure_s2f}

\refstepcounter{figure}
\label{figure_s3}
\refstepcounter{subfigure}
\label{figure_s3a}
\refstepcounter{subfigure}
\label{figure_s3b}
\refstepcounter{subfigure}
\label{figure_s3c}
\refstepcounter{subfigure}
\label{figure_s3d}
\refstepcounter{subfigure}
\label{figure_s3e}
\refstepcounter{subfigure}
\label{figure_s3f}

\refstepcounter{figure}
\label{figure_s4}
\refstepcounter{subfigure}
\label{figure_s4a}
\refstepcounter{subfigure}
\label{figure_s4b}
\refstepcounter{subfigure}
\label{figure_s4c}
\refstepcounter{subfigure}
\label{figure_s4d}
\refstepcounter{subfigure}
\label{figure_s4e}
\refstepcounter{subfigure}
\label{figure_s4f}

\refstepcounter{figure}
\label{figure_s5}
\refstepcounter{subfigure}
\label{figure_s5a}
\refstepcounter{subfigure}
\label{figure_s5b}
\refstepcounter{subfigure}
\label{figure_s5c}
\refstepcounter{subfigure}
\label{figure_s5d}
\refstepcounter{subfigure}
\label{figure_s5e}
\refstepcounter{subfigure}
\label{figure_s5f}

\refstepcounter{figure}
\label{figure_s6}
\refstepcounter{subfigure}
\label{figure_s6a}
\refstepcounter{subfigure}
\label{figure_s6b}
\refstepcounter{subfigure}
\label{figure_s6c}
\refstepcounter{subfigure}
\label{figure_s6d}
\refstepcounter{subfigure}
\label{figure_s6e}
\refstepcounter{subfigure}
\label{figure_s6f}

\refstepcounter{figure}
\label{figure_s7}
\refstepcounter{subfigure}
\label{figure_s7a}
\refstepcounter{subfigure}
\label{figure_s7b}
\refstepcounter{subfigure}
\label{figure_s7c}
\refstepcounter{subfigure}
\label{figure_s7d}
\refstepcounter{subfigure}
\label{figure_s7e}
\refstepcounter{subfigure}
\label{figure_s7f}

\end{document}